\documentclass[12]{iopart}

\expandafter\let\csname equation*\endcsname\relax
\expandafter\let\csname endequation*\endcsname\relax

\usepackage{graphicx}
\usepackage{amssymb}
\usepackage{amsmath}
\usepackage{bm}
\usepackage{url}
\usepackage{epstopdf}
\usepackage{iopams}

\begin{document}
\title[Electron Beam Dispersion in the FEL]{Velocity Dispersion of Correlated Energy Spread Electron Beams in the Free Electron Laser}

\author{L.T. Campbell$^{1,2}$ and A.R. Maier$^3$}
\address{1 SUPA, Department of Physics, University of Strathclyde, Glasgow, UK}
\address{2 ASTeC, STFC Daresbury Laboratory and Cockcroft Institute, Warrington, United Kingdom}
\address{3 Center for Free-Electron Laser Science and Department of Physics, \\ University of Hamburg, Luruper Chaussee 149, 22761 Hamburg, Germany}
\eads{\mailto{lawrence.campbell@strath.ac.uk}, \mailto{andreas.maier@desy.de}}
\submitto{\NJP}


\maketitle

\begin{abstract}
The effects of a correlated linear energy/velocity chirp in the electron beam in the FEL, and how to compensate for its effects by using an appropriate taper (or reverse-taper) of the undulator magnetic field, is well known. The theory, as described thus far, ignores velocity dispersion from the chirp in the undulator, taking the limit of a `small' chirp. In the following, the physics of compensating for chirp in the beam is revisited, including the effects of velocity dispersion, or beam compression or decompression, in the undulator. It is found that the limit of negligible velocity dispersion in the undulator is different from that previously identified as the small chirp limit, and is more significant than previously considered. The velocity dispersion requires a taper which is non-linear to properly compensate for the effects of the detuning, and also results in a varying peak current (end thus a varying gain length) over the length of the undulator. The results may be especially significant for plasma driven FELs and low energy linac driven FEL test facilities.
\end{abstract}



\section{Introduction}

The Free Electron Laser (FEL) is now established as the brightest source of coherent hard x-rays in the world, with facilities currently operational in the USA~\cite{lcls} and Japan~\cite{sacla}, and about to come online in Switzerland~\cite{swissfel} and Hamburg~\cite{exfel} in the near future. 

Owing to its flexibility and capacity for improvement, methods to further advance the radiation output are currently a topic of much research, and there are many methods proposed to improve the temporal coherence, increase the ouput power, and produce otherwise different types of output, such as multi-peaked spectra, pulse trains, X-rays with orbital angular momentum, or isolated short pulse output (see \cite{HBSASE}-\cite{OAMFEL}; see also \cite{natphot} and references therein).

There is also great interest in reducing the size and cost of future FELs by utilizing novel accelerator technology. Plasma accelerators are considered a promising future driver of FELs, with their high accelerating gradients and large peak currents. The electron beams typical of plasma accelerators possess small emittance, a large energy spread, and are very short compared to beams from more conventional linac sources. These characteristics provide challenges in beam transport both to and through the undulator. With regard to the FEL gain, the large energy spread is potentially the most deleterious feature at first glance, but measurements and simulations imply that a large proportion of the energy spread is corellated with the temporal bunch coordinate. 

This chirp in the beam energy causes a detuning in the FEL resonant frequency along the length of the bunch, and it is well known that this can be compensated for with an appropriate tapering of the undulator magnetic field~\cite{saldintap}. The energy difference between the front and the back of the electron bunch will result in a velocity dispersion as the beam propagates, but it has generally been assumed that this can be neglected in the FEL. Thus, previous analytic work describing the use of undulator field tapering to compensate the beam chirp neglected this consideration. 

However, with the increased interest in novel accelerator concepts as FEL drivers, \textit{e.g.} use of plasma accelerators \cite{PRX}-\cite{cite:huang2012} or the synthesis of broadband beams from linacs as in~\cite{BBD, felbbd}, the case of larger chirps has become more relevant. In this regime, dispersive effects can no longer be ignored, and the beam current and energy spread are a function of propagation distance through the undulator. Consequently, the gain length of the FEL is then itself a function of distance. In addition, dispersion due to the chirp will cause the gradient of the chirp to vary upon propagation, meaning that the taper necessary to compensate the chirp is also a function of undulator propagation length, and will not be linear.

FEL codes which employ `slices' with periodic boundaries to model the electron beam~\cite{MEDUSSA}-\cite{GINGER} cannot model this dispersion properly, as the electrons cannot travel between slices, and so cannot model any current redistribution through the undulator. The length of an individual slice is fixed and thus cannot model the disperions-induced locally varying decompression. In addition, the Slowly Varying Envelope Approximation (SVEA)~\cite{SVEA} means that they cannot model a broadband range of frequencies produced by large energy differences due to the chirp and/or a large taper. So-called `unaveraged' FEL codes \cite{POP}-\cite{maroli} are free of these limitations. For this reason, the unaveraged 3D FEL code Puffin~\cite{POP} is used in the following analysis.

In the following, the theory of compensating a beam chirp with an undulator taper is revisited, now including the effects of a velocity dispersion. The limits on when this dispersion is important are identified, and a more general expression for the taper required to compensate for an initially linear chirp is found. It is found that the limit of negligible dispersion is different, and more prohibitive, than the previously identified limit. The resulting beam compression or decompression varies the peak current, impacting the gain, even when cancelling the detuning effect with a taper. Using an initially linear chirp and compensating for the detuning with the new, correct taper, the effect on the gain from the beam compression is isolated from the detuning effect, and calculated analytically, and is compared to numerical simulations. It is shown that a linear energy chirp does not disperse linearly, and so in this case the detuning effect cannot be completely compensated for with a linear taper. Finally, 3D FEL simulations are presented to illustrate the effects on the output power of including significant velocity dispersion.

Note that typically the term `taper' refers to the technique of reducing the undulator magnetic field, and `reverse-taper' refers to the opposite; in the following, for brevity, we use the term `taper' in a more general sense as altering the magnetic field, either increasing or decreasing.

\section{Revisiting the Theory in Scaled Notation}

Using the scaled notation of~\cite{POP}, the propagation distance through the undulator is scaled to the 1D gain length, and the temporal coordinate in the stationary radiation frame is scaled to the 1D cooperation length, so that, respectively,
\begin{align}
\bar{z} = \frac{z}{L_g} \\
\bar{z}_2 = \frac{ct-z}{L_c}.
\end{align}

The scaled axial velocity of the j$^{th}$ electron is defined as
\begin{align}
p_{2j} = \frac{d \bar{z}_{2j}}{d\bar{z}} =  \frac{\beta_{zr}}{1 - \beta_{zr}} \frac{1-\beta_{zj}}{\beta_{zj}}  \label{p2eqn}
\end{align}
where $\beta_{zj} = v_{zj}/c$ is the $z$ velocity in the undulator normalised to the speed of light. The subscript $r$ denotes some reference velocity, which is usually sensible to take as the mean velocity of the beam, but which in general may be any velocity, as the model presented in~\cite{POP} allows a broadband description of both the radiation field and the electron energies. The resonant frequency corresponding to this reference velocity is then determined by
\begin{align}
k_r = \frac{\beta_{zr}}{1-\beta_{zr}} k_w,
\end{align}
and so, from eqn (\ref{p2eqn}), the electrons with $p_{2j}=1$ are resonant with this reference frequency.

Tapering is achieved by varying $\alpha(\bar{z}) = \bar{a}_w(\bar{z}) / \bar{a}_{w0}$, which is the relative change in the magnetic undulator field from its initial value, as defined in~\cite{2col}.

The gradient of an electron beam chirp may then be defined as 
\begin{align}
\frac{dp_2}{d\bar{z}_2}  \approx  - \frac{2}{\gamma_r} \frac{d \gamma}{d\bar{z}_2},
\end{align}
only when assuming small deviations in energy, assuming a small chirp so that
\begin{align}
\frac{dp_2}{d\bar{z}_2}  \ll 1,  \label{oldcond2}
\end{align}
and assuming small deviations in the undulator magnetic field, $\alpha \approx 1$.

Rewriting the formula for the taper required to compensate the detuning effect~\cite{saldintap} from a chirp in the above notation, we obtain
\begin{align}
\frac{d\alpha}{d\bar{z}} = - \frac{1 + \bar{a}_{w0}^2}{\bar{a}_{w0}^2}\frac{1}{\gamma_r}\frac{d \gamma}{d \bar{z}_2}  \label{oldtap}
\end{align}

\section{Dispersive and Broadband effects}

To take into account dispersive effects, it is convenient to describe the system using the $p_2$ phase space. $p_{2j}$ is the scaled velocity of the j$^{th}$ electron in the $\bar{z}_2$ frame, and so describes, linearly, how the beam will disperse. It also linearly measures the resonant wavelength of the electron; from eqn (\ref{p2eqn})
\begin{align}
p_{2j} = \frac{k_r}{k_j},
\end{align}
so it is the inverse of the frequency scaled to the reference frequency. 

\begin{figure}[t]
\centering
\includegraphics[width=100mm]{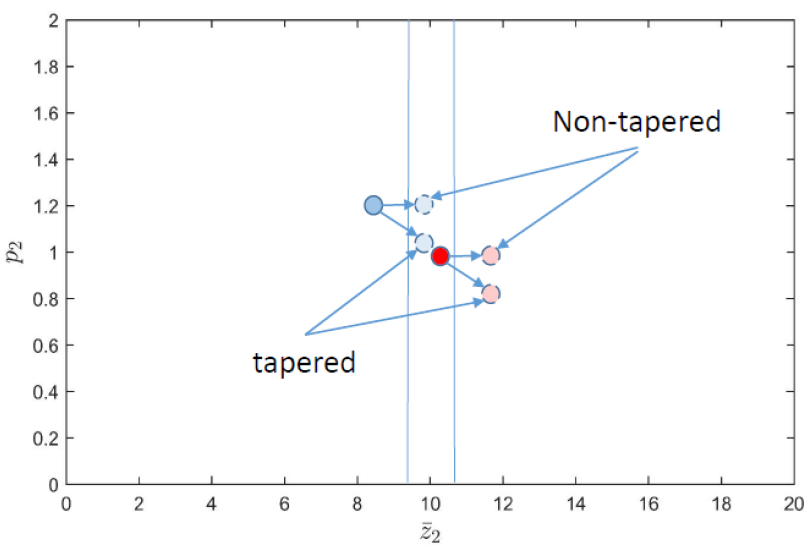}
\caption{Showing the manipulation of $p_2$ by variation of the undulator magnetic field $\alpha$. By altering the magnetic field, one may guide the blue electron to the correct value of $p_2$ to be resonant with the radiation in the slice indicated.}
\label{z2p2comp}
\end{figure}

Relaxing the constraint on the energies - once again allowing large energy changes - then eqn (\ref{oldtap}) is no longer correct. In the 1D limit, and using a helical wiggler, from equation (\ref{p2eqn}), $p_{2j}$ may be defined as a function of $\alpha$ and $\gamma$ as
\begin{align}
p_{2j}(\bar{z}) = \frac{\gamma_r^2}{\gamma_j^2} \Bigl(\frac{1 + \alpha(\bar{z})^2 \bar{a}_{w0}^2}{1 + \bar{a}_{w0}^2}\Bigr),  \label{p2falpha}
\end{align}
under the approximation that $\gamma_j, \gamma_r \gg 1$, ignoring any transverse velocity spread (1D limit), and ignoring any interaction with the radiation field (in the planar wiggler, one obtains the equivalent expression for $p_{2j}$ averaged over the wiggle motion). 

Using this definition, figure~\ref{z2p2comp} shows the effect of tapering in the $(\bar{z}_2, p_2)$ phase space, and shows what occurs when compensating for energy changes correlated in $\bar{z}_2$. The red electron, initially in the slice indicated, emits radiation at frequency $k_r$ before slipping back to the right. Recall this is the stationary radiation frame, and the head of the pulse is to the left. The blue electron, slipping back into the thin slice, finds itself interacting with radiation it is not resonant with. By varying, or tapering, the magnetic field $\alpha$, the value of $p_2$ of the blue electron can be manipulated, and reduced to the red electron's original value of $p_2$; therefore it is now resonant with the radiation in the slice originally emitted by the red electron. 

\begin{figure}[t]
\centering
\includegraphics[width=90mm]{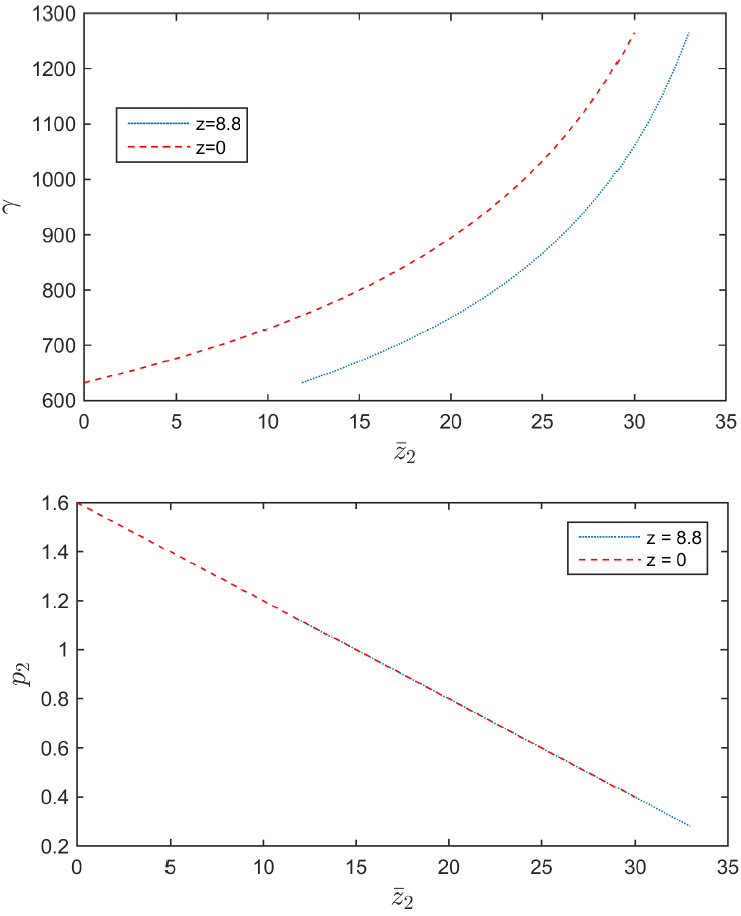}
\caption{Top: The electron beam mean energy $\gamma$ as a function of scaled temporal coordinate $\bar{z}_2$ at the start (red) and end (blue) of the undulator. Bottom: Same beam, now plotting the mean $p_2$ of the beam. The conversion from $p_2$ to $\gamma$ can be obtained from eqn (\ref{p2falpha}). This is the stationary radiation frame, and the head of the beam is to the left, so the beam slips backwards through the field from left to right.}
\label{linp2}
\end{figure}

Consequently, if an electron beam has an initial linear chirp in $p_2$, so that
\begin{align}
\frac{dp_2}{d\bar{z}_2}\Bigl|_{\bar{z}=0} = m,
\end{align}
then the correct magnetic field taper to ensure the beam stays resonant should cause each electron to follow the line of the chirp defined by $m$. Figure~\ref{linp2} shows this. It plots the mean energy of a beam, and the corresponding mean $p_2$, as a function of $\bar{z}_2$, at the start ($\bar{z}=0$) and end of an undulator tapered to compensate for the chirp. The taper may be derived from equations (\ref{p2falpha}) and (\ref{p2eqn}), forcing $dp_{2j} / d\bar{z} = m$ and $d\gamma_{j} / d\bar{z} = 0$, and solving for $\alpha$. The solution is found to be:
\begin{align}
\alpha = \frac{1}{\bar{a}_{w0}}\sqrt{ \exp(m\bar{z})(1 + \bar{a}_{w0}^2) - 1 },  \label{corralpha}
\end{align}  
which reduces to the solution of equation (\ref{oldtap}) only when 
\begin{align}
|m\bar{z}| \ll 1  \label{cond1}
\end{align}
and
\begin{align}
\frac{\bar{a}_{w0}^2}{1+\bar{a}_{w0}^2} \sim 1. \label{cond2}
\end{align}
For FEL's using magnetic undulators, where $\bar{a}_{w0}^2 \gtrsim 1$, condition (\ref{cond2}) is satisfied.

To measure the beam compression or decompression from this linear $p_2$ chirp, remembering that $p_2$ is the velocity of the electron in $\bar{z}_2$, then the change in the pulse width $\sigma_{z2}$ is
\begin{align}
\frac{d \sigma_{z2}}{d\bar{z}} = m \sigma_{z2}(\bar{z}),
\end{align}
so a stretch factor $S$ may be defined as
\begin{align}
S(\bar{z}) = \frac{\sigma_{z2}(\bar{z})}{\sigma_{z20}} = \exp(m\bar{z})
\end{align}
From this, it is seen that condition (\ref{cond1}) is the limit of negligible dispersion in the undulator. This is different from the limit of a small chirp as previously identified in eqn (\ref{oldcond2}), which is simply
\begin{align}
|m| \ll 1.  \label{smallm}
\end{align}
For a typical SASE FEL, the undulator is approximately $\bar{z} \approx 10-15$ long, so the dispersive condition at the end of the undulator is more restrictive than the previously defined `small chirp' limit by around an order of magnitude. For LPWA FELs, which have a larger slice energy spread and larger $\rho$, the undulator length may even be $\bar{z} \approx 30-35$ (recall $\bar{z}$ is scaled to the 1D gain length).

\begin{figure}[b]
\centering
\includegraphics[width=120mm]{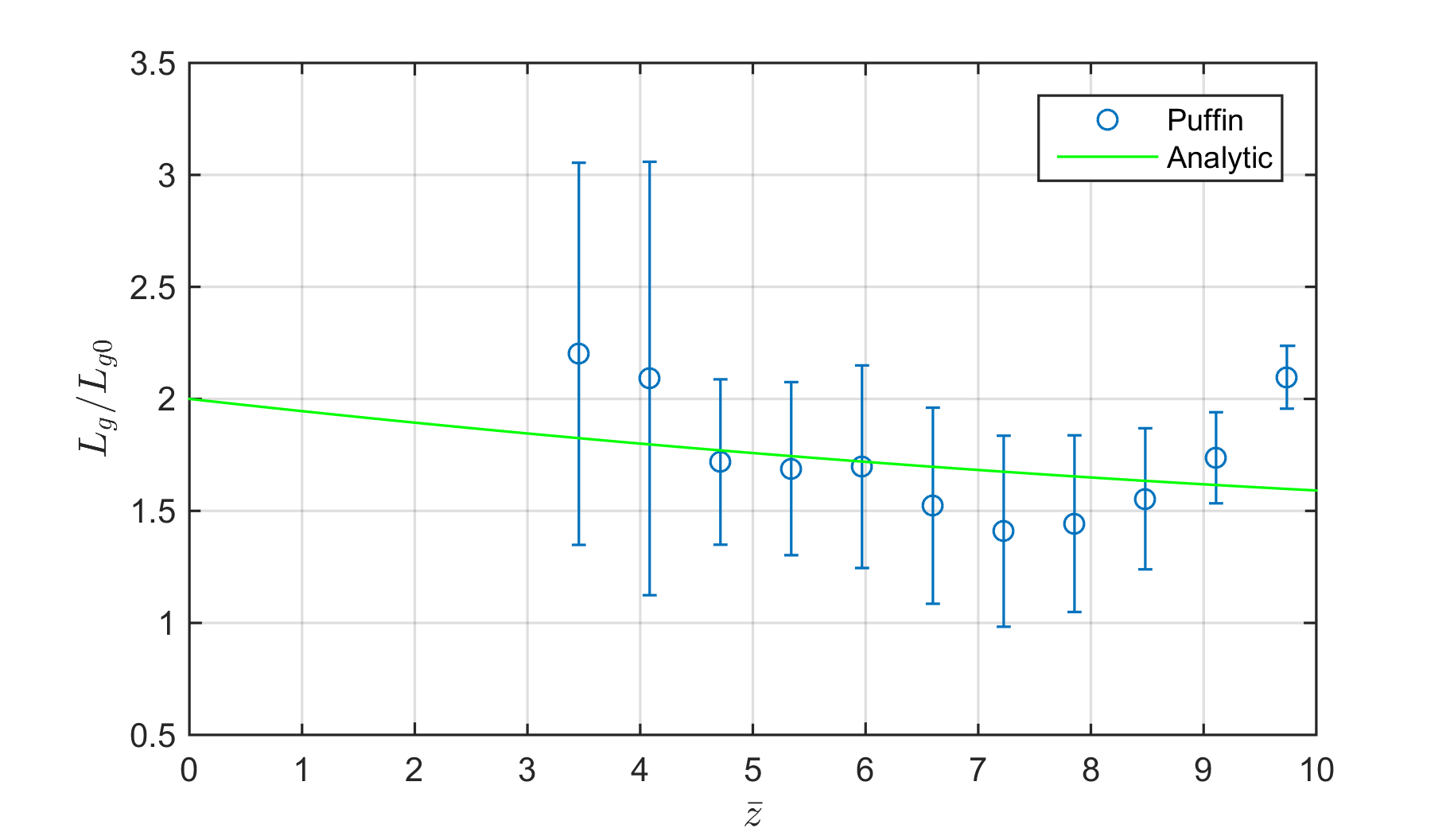}
\caption{Variation in gain length as a function of distance through the undulator due to dispersive effects. Analytic from eqn (\ref{anlg}) (green) compared to the gain length, measured from the numerical result produced in Puffin (blue, circles). The result from Puffin is averaged over 10 shots, with the error bars indicating the standard deviation of the result.}
\label{lgvar}
\end{figure}

\section{Measuring the Effect on the Gain Length}

The dispersion has an effect on the `3D' gain length~\cite{MXie}, as the compression/decompression will cause a change in the peak current and energy spread of the beam. The change in peak current can be analytically estimated very simply by
\begin{align}
I(\bar{z}) = \frac{I_0(\bar{z}=0)}{S(\bar{z})}.
\end{align}

The dispersion will also alter the localised, or `slice' energy spread of the beam. However, in this case, when using a linear chirp in $p_2$ with the taper in equation (\ref{corralpha}), every electron follows the line with gradient $m$ in the $(\bar{z}_2,p_2)$ phase space (see figure~\ref{linp2}), so the slice $p_2$ spread does not change despite the compression/decompression.  This corresponds to a variation in the peak transverse velocity, which is controlled by the peak magnetic field (it is, of course, this control of the magnitude of the transverse wiggle which allows one to control the resonant frequency of the FEL through the undulator magnetic field). 

\begin{figure}[t]
\centering
\includegraphics[width=90mm]{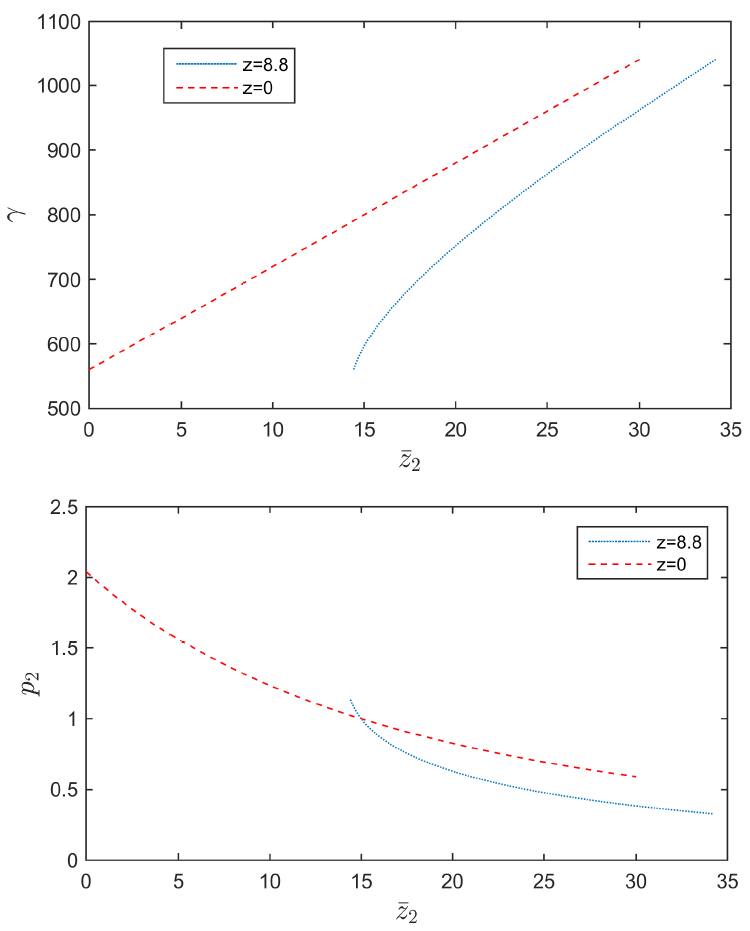}
\caption{Same as figure~\ref{linp2}, now using an initially linear chirp in energy $\gamma$. The undulator is now tapered to try to keep the mean electron beam $p_2$ constant at $\bar{z}_2 = 15$. In this case, so that it is resonant with the reference frequency $k_r$, $p_2=1$ at $\bar{z}_2=15$. The undulator taper is calcluated numerically, and is not linear.}
\label{lingam}
\end{figure}

The other consideration is that the gain length is different for each frequency; here, the frequency is linearly corellated with $\bar{z}_2$, and, because the taper is compensating perfectly, this correlation is fixed across the full undulator. Again refering to figure~\ref{linp2}, the mean $p_2$ at an instantaneous point in $\bar{z}_2$ remains constant, but the corresponding mean energy (from the top plot) is very different. Picking a coordinate initially in the center of the beam, $\bar{z}_{2c}$, with corresponding beam energy $\gamma_c$, which is a function of $\bar{z}$, then the normailised energy of the electron resonant with the fixed frequency is given by
\begin{align}
\Gamma = \frac{\gamma_c}{\gamma_{c0}} =  \Bigl(  \frac{1 + \alpha^2 \bar{a}_{w0}^2}{1 + \bar{a}_{w0}^2} \Bigr)^{1/2},
\end{align}
where $\gamma_{c0} = \gamma_{c}(\bar{z}=0)$.

From the definition of the FEL parameter, the gain length then varies as
\begin{align}
L_g(\bar{z}) = \frac{S(\bar{z})^{1/3} \Gamma(\bar{z})}{\alpha(\bar{z})^{2/3}} L_{g0} \label{anlg}
\end{align}
where $L_{g0}$ is the gain length at $\bar{z}=0$, and the gain length as referred to here is the M. Xie gain length, with only the energy spread parameter included.

A comparison of this analytic expression with the unaveraged FEL code Puffin is shown in figure~\ref{lgvar}. Relevant parameters used are $\rho = 0.01$, $\bar{a}_{w0} = 2$, $\gamma_r = 800$ and $m=-0.04$, and slice spread of $\sigma_\gamma/\gamma_r=1\%$. The gain length from Puffin is measured numerically from the radiated energy narrowly filtered around the frequency at $\bar{z}_{2c}$, and compares well with the analytic result. Note that the exponential gain region is $\bar{z} \approx 4$ to $\approx 8$; before this is the startup regime where there is no gain, and after this the system is in saturation. There is good agreement in the exponential gain regime.

By using a linear chirp in \textit{energy}, the beam compresses asymmetrically, and it is not possible the compensate for the detuning effect for all frequencies simultaneously. Figure \ref{lingam} plots the same quantities as figure \ref{linp2}, but with a linear energy chirp, and the taper is calculated numerically to keep the reference frequency at $\bar{z}_2=15$ interacting with electrons resonant with it (so, in this case, keeping $p_2=1$). The same can be done for any frequency emitted, so it is possible to preferentially compensate for certain frequencies, but it is not possible to properly compensate for all frequencies across the bunch simultaneously. 

However, this does not necessarly result in a higher power at that frequency. Other factors, such as the energy and slice spread, change differently for each frequency. Only the detuning effect is being compensated for; the other quantities (\textit{e.g.} current), varying asymmetrically across the bunch, may result in less or more gain at other frequencies when all effects are accounted for.


Consequently, there is a large range of tapers which can be considered `optimum'. But the detuning effect can only be completely removed across the whole bunch, for all frequencies, when the beam has a linear chirp in $p_2$, and using the taper described in equation (\ref{corralpha}). In that case, the effect on the gain length can be easily predicted.

Note that, in the above, only 1D effects have been taken into account. There is no examination of the change in diffraction parameter, beam divergence parameter \textit{etc} (from \cite{MXie}) occuring as a result of the dispersion. 

\section{3D LWA Example}

To illustrate the effects in a practical example, results from 3D Puffin simulations using parameters which may be expected from a laser plasma accelerator driven SASE FEL are now shown, displaying strong dispersive effects. The simulation uses a beam charge of $Q=100pC$, energy $E = 400MeV$, length of $\approx 6 \mu m$ in a flat-top current profile,  homogeneous or `slice' spread of $\sigma_\gamma / \gamma = 0.7 \%$, and normalised emittance of $1\mu m$. The beam is longer than might be expected from a `typical' LWA, to minimize the effects of a varying interaction length \cite{statstretch}, which would further complicate the behaviour. To keep the current similar to our generic LWA case, the total beam charge is consequently increased from a usual scenario. Note that in Puffin it is necessary to tail off the edges of the flat top current profile with a short gaussian profile to remove the effects of Self-Amplified CSE (SACSE) \cite{femfel}, leaving only the SASE process.

This beam is injected into a helical undulator with initial undulator parameters $\bar{a}_{w0}=2.0$, $\lambda_w=2.5cm$, giving a FEL parameter $\rho=0.0128$, a 1D gain length of $L_{g0} = 0.155m$, and resonant wavelength $\lambda_{r0} \approx 100$nm. 

The previous $1$D analysis described an idealised system with linear chirps in the scaled velocity $p_2$, and used non-linear tapers. In practice, non-linear tapers as described in equation (\ref{corralpha}) can be mechanically more difficult to realise than linear tapers, and the beam energy chirp is usually taken as linear. For this reason, in this example only linear {\it energy} chirps and magnetic field tapers are considered.

Two cases are considered; the first with no chirp and no taper, and the second with a linear energy chirp (such that the higher energies are towards the front of the bunch) and a linear compensating taper (such that $\bar{a}_w$ increases along the undulator). The chirp gives an energy change of $1.5\%$ over $1 \mu$m, giving $m= 0.02$ at the center of the bunch, so condition (\ref{smallm}) is perhaps satisfied; however, because the undulator is $\approx 26$ $1$D gain lengths long, condition (\ref{cond1}) is {\it not} satisfied, and the FEL interaction is strongly affected by the velocity dispersion.

3D effects here introduce more complicated effects on the gain, which include a varying betatron wavenumber with beam energy and a varying diffraction length with frequency. As a result, the effect on the gain is inherently more complex than in the 1D case. The beam radius is matched to the natural focusing channel of the undulator using the mean beam energy and the initial $\bar{a}_{w0}$, so that the whole beam has this same initial radius for both cases; but this may not be the optimal way to transversely match the beam in the chirped case.



The compensating linear tapers are scanned over a range of values to achieve an optimum output for the chirped case. As noted in the previous section, due to the chirp varying with the undulator distance in the case of a linear {\it energy} chirp, there is a range of linear tapers which appear to maximise the radiated energy. For example, here it was found that tapers of $d \alpha / d \bar{z} \approx 0.011 - 0.013$ all give extremely similar results in terms of the radiated energy. A value of $d \alpha / d \bar{z} = 0.012$ is used here.


\begin{figure}[t]
\centering
\includegraphics[width=130mm]{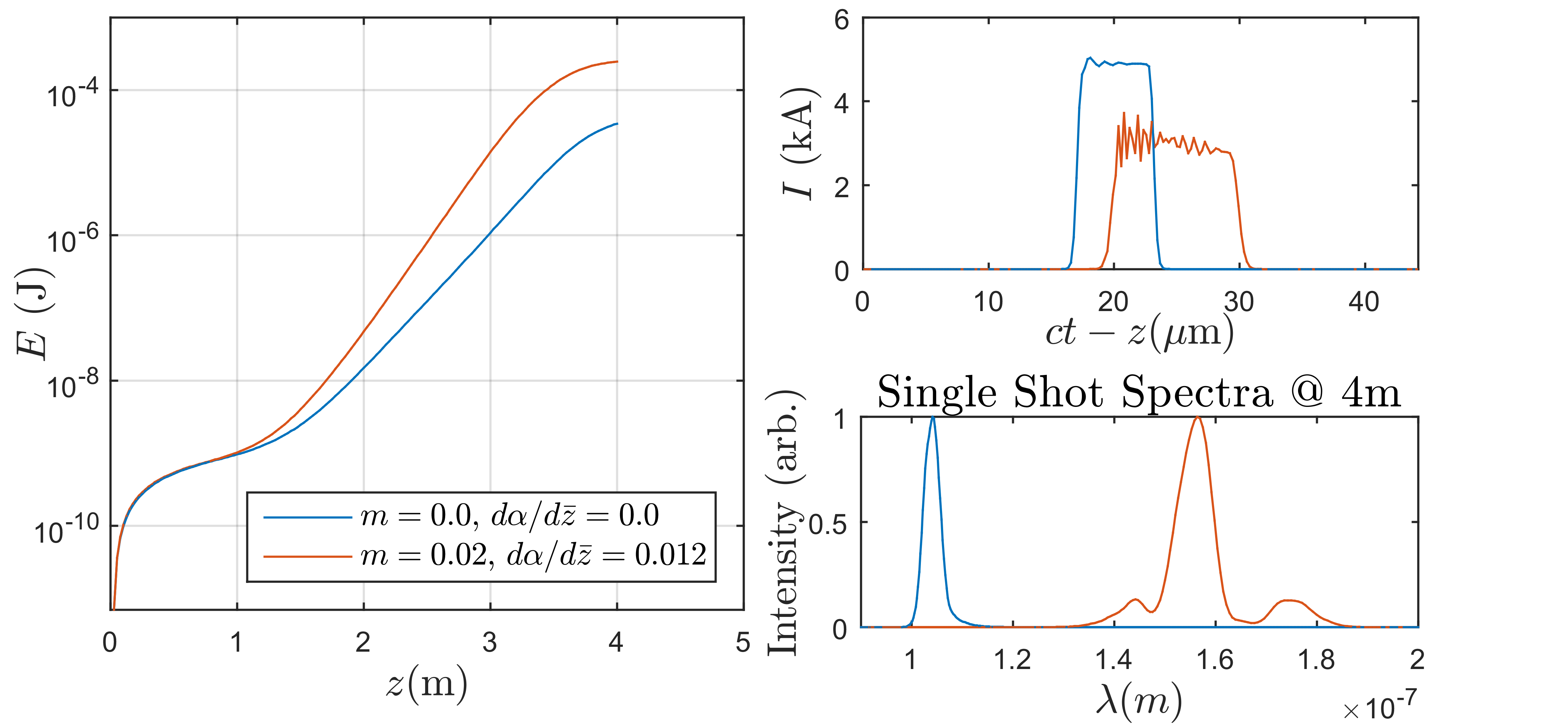}
\caption{Left: 3D Puffin simulations showing radiated energy for two different cases - a chirped and an unchirped beam for comparison. The undulator is linearly tapered to attempt to optimize the output in the chirped case. Each result is the average over 5 separate runs. Right, top: Current profile at the end of the undulator in each case. Right, bottom: On-axis intensity spectra at $4$m through the undulator, for a single shot of each case. Each case has been normalised to its own maximum value to easily see the relative wavelengths of the spectral peaks. }
\label{example1}
\end{figure}

Figure \ref{example1} shows the comparison between the chirped and unchirped cases. In both cases, the plotted result is the average over 5 shots. There is a significant difference in the energy output of over an order of magnitude, despite the undulator magnetic field taper used to minimize the detuning effect in the chirped case. In that case, where the beam decompresses, the peak current and slice spread are both reduced, producing an overall benefit to the gain. Note that depending on the initial slice energy spread of the beam, the opposite may be true: that compression aids the FEL process, and the decompression hinders it. 

The current profile at the undulator exit in each case is also plotted, showing the differences in the final beam profile. In addition to the different peak current and beam length, the mean positions of the beams are also different due to the extra slippage induced when changing the undulator magnetic field in the tapered case, implying a change in the mean amplified frequency.




The beam is short enough that the radiation slips from the tail to the head of the bunch more than once within the undulator, meaning the taper shifts the peak in the amplified frequencies -  a single frequency cannot be maintained at resonance for the full undulator distance. LWA generated beams are typically roughly an order of magnitude shorter than the beam used here, so this can be an expected feature whilst tapering to help an LWA beam lase. This changing frequency, due to the taper and the shortness of the bunch, further complicates the previous 1D analysis, which examined the gain at a {\it fixed} frequency. The spectra at the end of the $4$m undulator from a single shot from each case, also shown in figure \ref{example1}, exhibits this change in the amplified frequency in the chirped/tapered case, which is also implied by the different slippages from the plots of the current. One may see from eqn (\ref{oldtap}) that the relative frequency change when using linear tapers to compensate for an energy chirp will be reduced for larger $\bar{a}_{w0}$.  Also note that the bandwidth of amplified radiation in the chirped case is larger, as may be expected.

These 3D examples show the beam dispersion playing an important role in the FEL gain, and that this should be considered when measuring the gain if the expected beam chirp does not satisfy condition (\ref{cond1}), as it may not in the case of LWA's.

\section{Conclusion}

It has been shown that the beam dispersion in the undulator may be more important than previously acknowledged for presently achievable cases, especially in the realm of plasma driven FEL's, and the constraint on when it is relevant is actually $\sim$ an order of magnitude tighter than the previously identified condition of a `small' chirp. A simple model was presented to take the dispersion into account, which allows an analytic solution for a matched taper to eliminate the detuning effect, and allows one to isolate the effects of the dispersion and measure them. It is shown that the unaveraged code Puffin agrees with this result, in an extreme regime of FEL operation. Finally, an example shows the relevance of this work to laser plasma accelerator driven FEL's.

In the case of LWA's, a chicane may be used to stretch the beam before insertion into the unduator as in\cite{PRX} to control or diminish these effects, if they are deemed to be undesirable. If the beam disperses within the undulator, then this may need to be considered while optimizing the chicane parameters.

The discussion need not be limited to plasma accelerators, however, and the analysis presented here is entirely general, and can be applied to any case where a larger energy chirp is anticipated. In conventional linac driven UV/XRay FELs it is unlikely that the dispersion will be an issue at $>1$GeV, but for example at facilities used for a proof-of-principle demonstration of new FEL techniques at lower energy (\textit{e.g} the proposed UK CLARA facility \cite{CLARA} with energy $\sim 150-250$MeV), these effects may also be important.

\section{Acknowledgements}

We gratefully acknowledge the support of Science and Technology Facilities Council Agreement Number 4163192 Release \#3; ARCHIE-WeSt High Performance Computer, EPSRC Grant EP/K000586/1; EPSRC Grant EP/M011607/1; and John von Neumann Institute for Computing (NIC) on JUROPA at J\"ulich Supercomputing Centre (JSC), under project HHH20.

\end{document}